\documentclass[aip,jap,reprint,amsmath,amssymb,showpacs,showkeys,superscriptaddress]{revtex4-1}

\usepackage{graphicx}% Include figure files
\usepackage{dcolumn}% Align table columns on decimal point
\usepackage{bm}% bold math
\usepackage[mathlines]{lineno}% Enable numbering of text and display math
\begin{document}

%\preprint{PRB/v04_15.05.2013}

\title{Magnetoresistance control in granular Zn$_{1\textrm{-}x\textrm{-}y}$Cd$_{x}$Mn$_{y}$GeAs$_{2}$ nanocomposite ferromagnetic semiconductors}

\author{L.~Kilanski}
 \email{kilan@ifpan.edu.pl}
\affiliation{Institute of Physics, Polish Academy of Sciences, Al. Lotnikow 32/46, 02-668 Warsaw, Poland}

\author{I.~V.~Fedorchenko}
\affiliation{Kurnakov Institute of General and Inorganic Chemistry RAS, 119991 Moscow, Russia}
\affiliation{Lappeenranta University of Technology, P.O. Box 20, FI-53851 Lappeenranta, Finland}

\author{M.~G\'{o}rska}
\author{A.~\'Slawska-Waniewska}
\author{N.~Nedelko}
\author{A.~Podg\'{o}rni}
\author{A.~Avdonin}
\affiliation{Institute of Physics, Polish Academy of Sciences, Al. Lotnikow 32/46, 02-668 Warsaw, Poland}

\author{E.~L\"{a}hderanta}
\affiliation{Lappeenranta University of Technology, P.O. Box 20, FI-53851 Lappeenranta, Finland}

\author{W.~Dobrowolski}
\affiliation{Institute of Physics, Polish Academy of Sciences, Al. Lotnikow 32/46, 02-668 Warsaw, Poland}

\author{A.N.~Aronov}
\author{S.~F.~Marenkin}
\affiliation{Kurnakov Institute of General and Inorganic Chemistry RAS, 119991 Moscow, Russia}

\date{\today}

\begin{abstract}

We present studies of structural, magnetic and electrical properties of Zn$_{1\textrm{-}x\textrm{-}y}$Cd$_{x}$Mn$_{y}$GeAs$_{2}$  nanocomposite ferromagnetic semiconductor samples with changeable chemical composition. The presence of MnAs clusters induces in the studied alloy room temperature ferromagnetism with the Curie temperature, $T_{C}$, around 305$\;$K. The chemical composition of the chalcopyrite matrix controls the geometrical parameters of the clusters inducing different magnetoresistance effects in the crystals. The presence of ferromagnetic clusters in the alloy induces either negative or positive magnetoresistance with different values. The Cd-content allows a change of magnetoresistance sign in our samples from negative (for $x$$\,$$\approx$$\,$0.85) to positive (for $x$$\,$$\approx$$\,$0.12). The negative magnetoresistance present in the samples with $x$$\,$$\approx$$\,$0.85 is observed at temperatures $T$$\,$$<$$\,$25$\;$K with maximum values of about -32\% at $T$$\,$$=$$\,$1.4$\;$K and $B$$\,$$=$$\,$13$\;$T, strongly depending on the Mn content, $y$. The positive magnetoresistance present in the samples with $x$$\,$$\approx$$\,$0.12 is observed with maximum values not exceeding 50\% at $B$$\,$$=$$\,$13$\;$T and $T$$\,$$=$$\,$4.3$\;$K, changing with the Mn content, $y$.

\end{abstract}



% Transition-metal compounds, electrical conductivity of, 72.80.Ga
% Magnetic impurity interactions, 75.30.Hx
% Exchange interactions, 75.30.Et
% Magnetic semiconductors, 75.50.Pp

%\end{frontmatter}

\maketitle

\section{Introduction}

Complex ferromagnetic semiconductors (FMS) offer several advantages over the well known and intensively studied Mn-doped III-V, IV-VI, and II-VI materials.\cite{Kossut1993a, Dobrowolski2003a} FMS systems showing room temperature ferromagnetism are a subject of considerable interest in the recent years since they offer potential for being applied in spintronics. However, most of the literature reports about the magnetic properties of DMS systems show the Curie temperatures much lower than 300$\;$K, which makes these compounds of little use for practical applications. \\ \indent The nonexistence of room temperature FMS systems creates the need for the development of new compounds fulfilling technological requirements. Homogeneous FMS systems do not show the Curie temperature, $T_{C}$, with values exceeding 200$\;$K.\cite{Dietl2010a} Inhomogeneous FMS systems are more prospective from the point of view of their possible applications. Among many complex nanocomposite FMS the II-IV-V$_{2}$ chalcopyrite semiconductors doped with transition metal ions are recently intensively studied and developed.\cite{Erwin2004a, Picozzi2004a} It has been proven that the short-range magnetic interactions connected with the presence of magnetic clusters are responsible for high-temperature ferromagnetism in these alloys.\cite{Kilanski2009a, Kilanski2009b, Kilanski2010a} Large solubility of Mn ions in both Zn$_{1\textrm{-}x}$Mn$_{x}$GeAs$_{2}$ (Refs.$\;$\onlinecite{Kilanski2011a} and \onlinecite{Kilanski2013a}) and Cd$_{1\textrm{-}x}$Mn$_{x}$GeAs$_{2}$ (Ref.$\;$\onlinecite{Kilanski2014a}) crystals grown under thermodynamical equilibrium conditions and a rather large value of the Mn-ion conducting hole magnetic exchange constant, $J_{pd}$$\,$$\approx$$\,$0.75$\pm$0.09$\;$eV, have been observed indicating that the long range itinerant ferromagnetism is possible to obtain in this group of semiconductors. \\ \indent In the present paper we describe the problem of control of the magnetic and magnetotransport properties of nanocomposite ferromagnetic semiconductors based on the II-IV-V$_{2}$ chalcopyrite crystals alloyed with manganese ions. The control of the properties of the clusters is related to the chemical composition of the semiconductor matrix allowing the change of the physical mechanism of the magnetoresistance effects in the alloy. The possibilities of controlling the sign and the absolute value of the magnetoresistance via changes of the main physical mechanism responsible for the magnetoresistance will be presented for the Zn$_{1\textrm{-}x\textrm{-}y}$Cd$_{x}$Mn$_{y}$GeAs$_{2}$ crystals with different chemical compositions.

\section{Sample preparation and structural characterization}

\begin{figure}
 \includegraphics[width = 0.5\textwidth, bb = 5 15 540 585]
{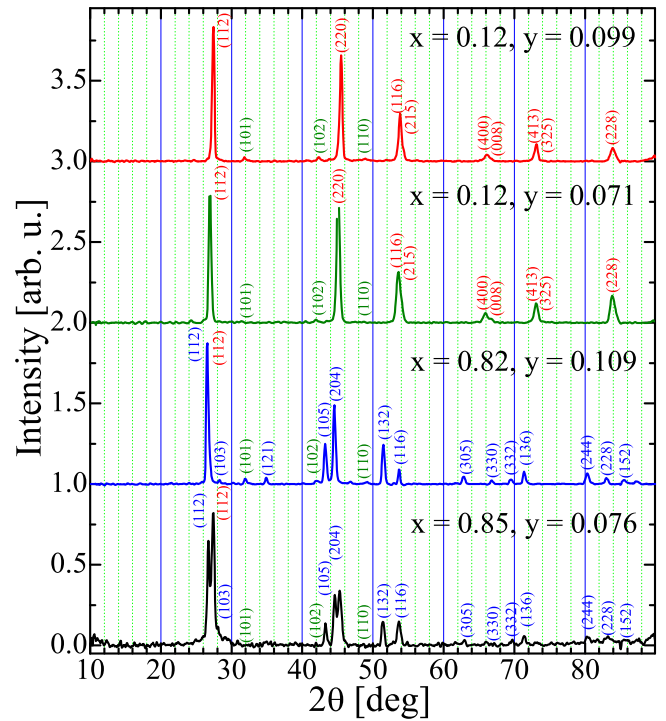}%
 \caption{\label{FigXRD} The XRD patterns for the selected Zn$_{1\textrm{-}x}$Cd$_{x}$GeAs$_{2}$:MnAs samples with different chemical compositions. Peaks in the diffraction patterns are marked with \emph{hkl} values for the known chalcopyrite ZnGeAs$_{2}$ (blue), CdGeAs$_{2}$ (red), and MnAs (green) phases.}
 \end{figure}

For the purpose of the current research two series (Zn- and Cd-rich) of Zn$_{1\textrm{-}x\textrm{-}y}$Cd$_{x}$Mn$_{y}$GeAs$_{2}$ samples were grown with the use of the direct fusion method. The sample synthesis was done in a few steps using as starting materials the high purity As (99,999\%), Zn, Cd, and Ge (99,99\%) single crystals. The synthesis procedure of the Zn$_{1\textrm{-}x\textrm{-}y}$Cd$_{x}$Mn$_{y}$GeAs$_{2}$ alloy is an extension of the procedure used in Ref.$\;$\onlinecite{Fedorchenko2014a} for the synthesis of Zn$_{1\textrm{-}x}$Cd$_{x}$GeAs$_{2}$ alloy. At first, the binary ZnAs$_{2}$ and CdAs$_{2}$ compounds were synthesized by direct interaction of As with Zn and Cd. The synthesis of the ZnAs$_{2}$ and CdAs$_{2}$ compounds prior to the synthesis of complex compounds was done in order to avoid high pressure processes involving arsenic that would occur during the synthesis of the Zn$_{1\textrm{-}x\textrm{-}y}$Cd$_{x}$Mn$_{y}$GeAs$_{2}$ solid solution. The structural quality of the as-grown ZnAs$_{2}$ and CdAs$_{2}$ crystals was checked by using the X-ray diffraction (XRD) and microstructure analysis. The analysis shows the successful synthesis of pure, single phased ZnAs$_{2}$ and CdAs$_{2}$ crystals, being precursors for the next synthesis steps. The second step of the synthesis procedure aimed at the synthesis of ternary ZnGeAs$_{2}$ and CdGeAs$_{2}$ compounds by synthesis of pure Ge with ZnAs$_{2}$ and CdAs$_{2}$ compounds. The structural quality of the fabricated ternary compounds was also checked with XRD and microstructure analysis. For the next step of the synthesis procedure only single phase crystals were used. The last part of the Zn$_{1\textrm{-}x\textrm{-}y}$Cd$_{x}$Mn$_{y}$GeAs$_{2}$ alloy synthesis was done with ZnGeAs$_{2}$, CdGeAs$_{2}$, and MnAs compounds, weighed in stoichiometric ratios, placed in the quartz ampoules (covered with a thin film of graphite to avoid the interaction of the synthesizing material with the ampoule). The ampoules with the synthesis material were evacuated to the pressure lower than 10$^{-6}$ mbar and sealed. The growth process was done at the temperature 1150$\;$K kept constant for 24 hours for homogenization of the crystal melt. The crystal melt was cooled down from 1150$\;$K to about 250$\;$K with the speed of about 150$\;$K/s. High cooling speed minimizes the size of MnAs inclusions in the as-grown crystals. \\ \indent The chemical composition of our samples was determined with the use of the energy dispersive x-ray fluorescence method (EDXRF). The measurements were done with the use of the Tracor X-ray Spectrace 5000 spectrometer. The as grown ingots were cut into thin slices (typically around 1$\;$mm thick) with the use of a precision wire saw. The maximum relative errors of the EDXRF technique do not exceed 10\% of the calculated values of the Cd and Mn content, $x$ and $y$, respectively. The results of the EDXRF measurements are gathered in Table.$\;$\ref{tab:MagnParams1}. As expected from the technological proportions of the alloying elements used for the synthesis the EDXRF data indicate that two groups of samples were obtained: (i) Zn-rich with $x$$\,$$\approx$$\,$0.12$\pm$0.01 and Cd-rich with $x$$\,$$\approx$$\,$0.85$\pm$0.09. The $x$ value for each of the group is almost constant within the uncertainty of the EDXRF method. The reason for the selection of Cd content $x$ lies in the phase diagram of ZnGeAs$_{2}$-CdGeAs$_{2}$, investigated in detail in Ref.$\;$\onlinecite{Fedorchenko2014a}. The solid solutions are present in all range of concentrations above 440$^{\textrm{o}}$C. Below this temperature the decay region of solid solution of Zn$_{1\textrm{-}x}$Cd$_{x}$GeAs$_{2}$ is observed in the range of Cd concentrations $x$ from 0.19 to 0.92 at room temperature. It is then not possible to synthesize crystals with $x$ in the range from 0.19 to 0.92. Therefore, we focused our studies over the boundary Cd content, at which Zn$_{1\textrm{-}x}$Cd$_{x}$GeAs$_{2}$ exists. All our samples belong to the boundary region, taking into account the uncertainty in the determination of the $x$ and $y$ values with the use of the EDXRF method and the fact, that the addition of Mn to the alloy may change the region at which the Zn$_{1\textrm{-}x\textrm{-}y}$Cd$_{x}$Mn$_{y}$GeAs$_{2}$ alloy exists. The main purpose of the synthesis procedure was to obtain samples with different Mn content, $y$. The data gathered in Table$\;$\ref{tab:MagnParams1} indicate that $y$ changes from 0.033 to 0.109 allowing a large change of the concentration of the MnAs clusters in the material. \\ \indent The high resolution X-ray diffraction measurements were performed in all our samples at room temperature using the multipurpose diffractometer. The Cu K$_{\alpha 1}$ x-ray radiation with the wavelength 1.5406$\textrm{\AA}$ was used. The obtained HRXRD patterns for our Zn$_{1\textrm{-}x\textrm{-}y}$Cd$_{x}$Mn$_{y}$GeAs$_{2}$ samples are shown in Fig.$\;$\ref{FigXRD}.
The XRD data indicate the presence of 3 phases with different concentrations, depending on the alloy composition. One is identified as a chalcopyrite structure with lattice parameters similar to the values reported for the ZnGeAs$_{2}$ compound equal to $a$$\,$$=$$\,$5.67$\;$$\textrm{\AA}$ and $c$$\,$$=$$\,$11.153$\;$$\textrm{\AA}$ (Ref.$\;$\onlinecite{Pfister1958a}). The second phase is identified as a chalcopyrite structure with the lattice parameters having values close to the literature data available for CdGeAs$_{2}$ compound, i.e. $a$$\,$$\approx$$\,$5.942$\;$$\textrm{\AA}$ and $c$$\,$$\approx$$\,$11.224$\;$$\textrm{\AA}$ (Ref.$\;$\onlinecite{Pfister1958a}). The third phase present in the XRD data is identified as MnAs phase. It was not possible to determine the crystal structure and the lattice parameters of the MnAs from the XRD data since MnAs contributes very little to the diffraction patterns. However, since both the hexagonal and the orthorhombic MnAs phase have significantly different magnetic properties the magnetometric methods will prove useful for the identification of the type of the MnAs inclusions present in our samples.

\section{Magnetic properties}

\begin{figure}
 \includegraphics[width = 0.45\textwidth, bb = 5 10 597 540]
{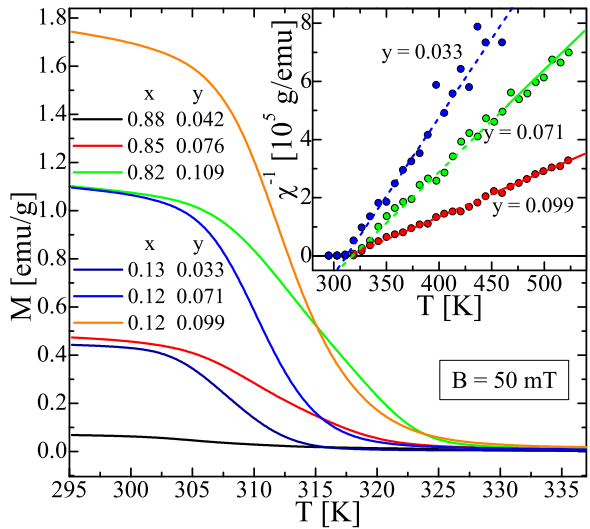}%
 \caption{\label{FigMTfc} Magnetization as a function of temperature for the Zn$_{1\textrm{-}x\textrm{-}y}$Cd$_{x}$Mn$_{y}$GeAs$_{2}$ crystals with different chemical compositions. The inset shows the inverse of the experimental magnetic susceptibility for samples with $x$$\,$$\approx$$\,$0.12 and 0.13 (points) and the curves fitted with the Curie-Weiss law.}
 \end{figure}

\begin{figure}
 \includegraphics[width = 0.5\textwidth, bb = 0 20 832 590]
{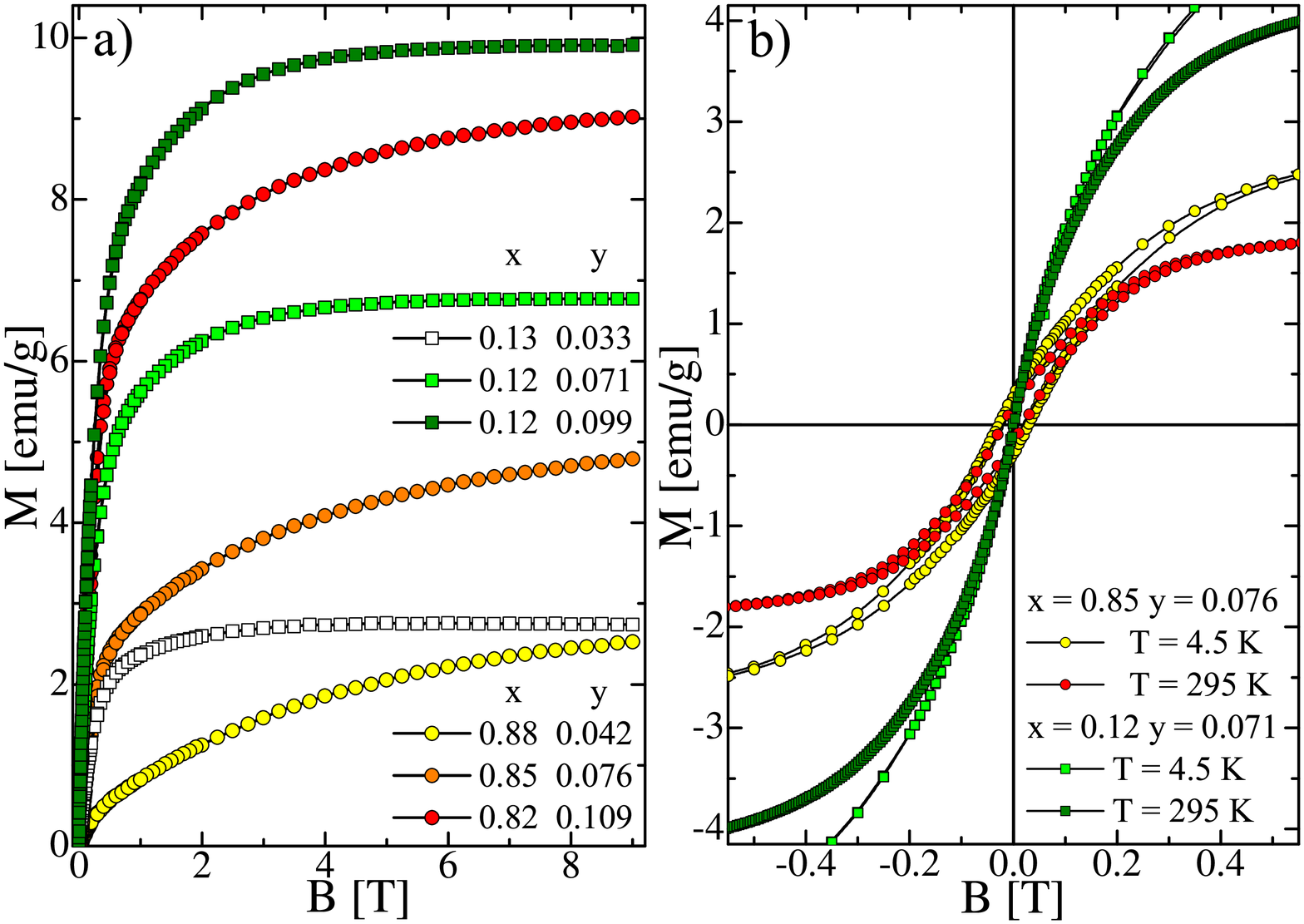}%
 \caption{\label{FigMvsB} Magnetization as a function of magnetic field obtained at different temperatures for the selected Zn$_{1\textrm{-}x\textrm{-}y}$Cd$_{x}$Mn$_{y}$GeAs$_{2}$ crystals with different chemical compositions.}
 \end{figure}

\noindent Magnetic properties of our Zn$_{1\textrm{-}x\textrm{-}y}$Cd$_{x}$Mn$_{y}$GeAs$_{2}$ crystals were studied with the use of two static magnetization, $M$, measurement techniques. The low-field magnetization measurements were performed with the use of the vibrating sample magnetometer (VSM) system. The measurements of the high-field static magnetization were performed with the use of the LakeShore 7229 magnetometer system. \\ \indent The temperature dependence of the magnetic moment was measured with the sample put in a static magnetic field of induction $B$$\,$$=$$\,$50$\;$mT over the temperature range from 270 to 550$\;$K. The results of the measurements for all our samples are presented in Fig.$\;$\ref{FigMTfc}.
The shape of the $M$($T$) curves indicates the presence of the magnetic phase transition at around 310$\;$K. It is a signature, previously observed for both Zn$_{1\textrm{-}x}$Mn$_{x}$GeAs$_{2}$ (Ref.$\;$\onlinecite{Kilanski2010a}) and Cd$_{1\textrm{-}x}$Mn$_{x}$GeAs$_{2}$ (Ref.$\;$\onlinecite{Kilanski2011b}), of the clustering of paramagnetic impurities into ferromagnetic MnAs clusters. The Curie temperature, $T_{C}$, of the MnAs inclusions present in our samples can be calculated from the inflection point of the $M$($T$) curve, i.e. where the condition
$\partial^{2}M/\partial T^{2}$$\,$$\rightarrow$$\,$0 occurs. The obtained values of $T_{C}$ are gathered in Table$\;$\ref{tab:MagnParams1}.
\begin{table*}[t]
\caption{\label{tab:MagnParams1}
Magnetization parameters obtained for the Zn$_{1\textrm{-}x\textrm{-}y}$Cd$_{x}$Mn$_{y}$GeAs$_{2}$ crystals with different chemical compositions.}
\begin{tabular}{cccccccc}
\hline \hline \smallskip
 $x$$\pm$$\Delta$$x$ & $y$$\pm$$\Delta$$y$ & $T_{C}$$\pm$$\Delta$$T_{C}$  &  $C$$\pm$$\Delta$$C$ (10$^{-4}$) & $\bar{y}_{\theta}$$\pm$$\Delta$$\bar{y}_{\theta}$ &  $\theta$$\pm$$\Delta$$\theta$  & $M_{S}$$\pm$$\Delta$$M_{S}$ & $\bar{y}_{m}$$\pm$$\Delta$$\bar{y}_{m}$  \\
 & & [K] &  [emu$\cdot$K/g] & & [K] & [emu/g]  &   \\ \hline
 0.88$\pm$0.08  & 0.042$\pm$0.004 & 306$\pm$1 & 2.0$\pm$0.1 & 0.015$\pm$0.001 & 312$\pm$1  & - & - \\
 0.85$\pm$0.08  & 0.076$\pm$0.007 & 310$\pm$1 & 7.4$\pm$0.4 & 0.055$\pm$0.003 & 309$\pm$1  & - & - \\
 0.82$\pm$0.08  & 0.109$\pm$0.01  & 313$\pm$1  & 8.7$\pm$0.3 & 0.065$\pm$0.002 & 310$\pm$1 & - & - \\ \hline
 0.13$\pm$0.01  & 0.033$\pm$0.003 & 308$\pm$1 & 1.7$\pm$0.1 & 0.011$\pm$0.001 & 310$\pm$1  & 2.7$\pm$0.1 & 0.028$\pm$0.003 \\
 0.12$\pm$0.01  & 0.071$\pm$0.007 & 310$\pm$1 & 3.4$\pm$0.1 & 0.023$\pm$0.001 & 307$\pm$1  & 6.8$\pm$0.2 & 0.071$\pm$0.007 \\
 0.12$\pm$0.01  & 0.099$\pm$0.009 & 312$\pm$1 & 7.1$\pm$0.2 & 0.047$\pm$0.002 & 309$\pm$1  & 9.9$\pm$0.3 & 0.10$\pm$0.01 \\  \hline \hline
\end{tabular}
\end{table*}
The values of $T_{C}$ do not change significantly with the amount of Mn ions in the material. However, for the samples with both low ($x$$\,$$\approx$$\,$0.12) and high ($x$$\,$$\approx$$\,$0.85) Cd content a slight increase of the $T_{C}$ with $y$ can be observed. It might be a signature of slight changes in the structural properties of the MnAs inclusions present in the material, i.e. changes in the stoichiometry and the lattice parameters. \\ \indent The magnetic susceptibility, $\chi_{dc}$$\,$$=$$\,$$\frac{\partial M}{\partial H}$$\big{|}_{T=const.}$ can be calculated using the magnetization data at selected constant temperatures and for $B$ values small enough to stay in the linear regime of $M$($B$) curves. The inverse of the magnetic susceptibility (plotted in the inset to Fig.$\;$\ref{FigMTfc}) at temperatures well above $T_{C}$, i.e. in a paramagnetic region should follow from the Curie-Weiss law expressed with the use of the following equation:
\begin{equation}\label{EqCWLaw}
    \chi_{dc} = \frac{C}{T - \theta} + \chi_{dia},
\end{equation}
where
\begin{equation}\label{EqCWLawCConst}
    C= \frac{N_{0} g^{2} \mu_{B}^{2} S(S+1) {\bar y_{\theta}}}{3k_{B}}.
\end{equation}
Here $C$ is the Curie constant, $\chi_{dia}$ is the diamagnetic contribution to the magnetic susceptibility originating from the host lattice, $N_{0}$ is the number of cation sites per gram, $g$ is the g-factor of the magnetic ion (for Mn $g$$\,$$=$$\,$2), $S$$\,$$=$$\,$5/2 is the spin-magnetic momentum of the Mn ion, $\mu_{B}$ is the Bohr magneton, $k_{B}$ is the Boltzmann constant, and ${\bar y_{\theta}}$ is the effective content of magnetically-active Mn. The experimental data obtained over the temperature range from 400$\;$K to 550$\;$K were fitted to the Eq.$\;$\ref{EqCWLaw} assuming the constant value of the diamagnetic contribution to the magnetic susceptibility. We took the $\chi_{dia}$ value estimated for ZnGeAs$_{2}$ equal to $\chi_{dia}$$\,$$=$$\,$$-2\times$10$^{-7}$$\;$emu/g (Ref.$\;$\onlinecite{Kilanski2013a}) for fitting the $M$($T$) dependence for the samples with low Cd-content, $x$$\,$$\approx$$\,$0.12, and $\chi_{dia}$$\,$$=$$\,$$-2.5\times$10$^{-7}$$\;$emu/g estimated in CdGeAs$_{2}$ (Ref.$\;$\onlinecite{Kilanski2014a}) for fitting the $M$($T$) dependence for the samples with high Cd-content, $x$$\,$$\approx$$\,$0.85. We fitted the experimental (Re($\chi_{AC}$))$^{-1}$($T$) curves with two fitting parameters: the Curie-Weiss temperature $\theta$ and the Curie constant $C$. The fitted curves are presented together with the experimental data in Fig.$\;$\ref{FigMTfc}. As we can clearly see the magnetic susceptibility of our samples can be very well reproduced with the use of the Curie-Weiss law defined with Eq.$\;$\ref{EqCWLaw}. The fitting parameters, $\theta$ and $C$, are gathered for all our samples in Table$\;$\ref{tab:MagnParams1}. The obtained values of $C$ can be used to calculate the amount of magnetically active Mn-ions, ${\bar y_{\theta}}$, using Eq.$\;$\ref{EqCWLawCConst}. For all our samples ${\bar y_{\theta}}$$\,$$<$$\,$$y$ indicating that a large fraction of the Mn ions present in the material is either magnetically inactive or has a charge state different from the Mn$^{2+}$ high-spin-state with the total magnetic momentum, $J$$\,$$=$$\,$$S$$\,$$=$$\,$5/2. The second fitting parameter obtained from fits to the Eq.$\;$\ref{EqCWLaw} is the Curie-Weiss temperature, $\theta$. The values of $\theta$ are close to $T_{C}$ indicating a lack of strong magnetic frustration in our samples, that could lead to significant difference between $\theta$ and $T_{C}$. \\ \indent The isothermal magnetic field dependence of the magnetization was measured for the samples with different chemical compositions. Two series of the $M$($B$) curves were obtained: (i) at $T$$\,$$=$$\,$4.5$\;$K the magnetization curves were measured using the Weiss extraction method employed into the LakeShore 7229 magnetometer system in the magnetic field up to $B$$\,$$=$$\,$9$\;$T and (ii) at $T$$\,$$\approx$$\,$300$\;$K the magnetization curves were measured with the use of the VSM magnetometer system for the magnetic field up to $B$$\,$$=$$\,$1.5$\;$T. The two instruments had to be used to cover the large temperature range where the samples are ferromagnetic. The experimental results were corrected by subtracting the contribution of the sample holder. Because the $M$($B$) curves at low temperatures carry most information about the magnetic ions and for clarity of presentation only the results obtained at the lowest temperature, $T$$\,$$=$$\,$4.3$\;$K, and at room temperature are presented and analyzed. Examples of $M$($B$) curves obtained for the selected Zn$_{1\textrm{-}x\textrm{-}y}$Cd$_{x}$Mn$_{y}$GeAs$_{2}$ samples with different chemical compositions are presented in Fig.$\;$\ref{FigMvsB}.
The $M$($B$) curves for the samples with $x$$\,$$\approx$$\,$0.12 reach saturation at moderate magnetic fields, $B$$\,$$<$$\,$9$\;$T, used in our experiments. The saturation magnetization, $M_{S}$, the quantity important for the estimation of the effective Mn content in our samples, for the samples with $x$$\,$$\approx$$\,$0.12 is gathered in Table.$\;$\ref{tab:MagnParams1}. However, for the samples with $x$$\,$$\approx$$\,$0.8 the $M$($B$) curves do not show saturation. The nonsaturating behavior of the $M$($B$) curves originates from the higher level of disorder present in the samples with $x$$\,$$\approx$$\,$0.8 introducing to the system antiferromagnetic interactions that lead to magnetic frustration. The saturation magnetization of the system with magnetic impurities can be described with the use of the following equation
\begin{equation}\label{EqMSx}
    M_{S} = \bar{y}_{m} N_{0} \mu_{B} g S,
\end{equation}
where $B_{S}$ is the Brillouin function and $\bar{y}_{m}$ is the effective content of magnetically active Mn. The obtained values of $\bar{y}_{m}$ gathered in Table$\;$\ref{tab:MagnParams1} show that $\bar{y}_{m}$$\,$$\approx$$\,$$y$.   \\ \indent The magnetization of our samples shows the presence of magnetic hysteresis only for the samples with high Cd-content, $x$$\,$$\approx$$\,$0.85. The different behavior of the magnetic hysteresis indicates a sharp change of the structural parameters of the MnAs clusters. The coercive field and remnant magnetization, $H_{C}$ and $M_{R}$, respectively, change with the amount of Mn ions in the samples with $x$$\,$$\approx$$\,$0.85. Both values are increasing functions of the Mn content, $y$, changing from $H_{C}$$\,$$=$$\,$18$\;$mT and $M_{R}$$\,$$=$$\,$0.033$\;$emu/g to $H_{C}$$\,$$=$$\,$28$\;$mT and $M_{R}$$\,$$=$$\,$0.39$\;$emu/g, for the samples with $y$$\,$$=$$\,$0.042 and 0.109, respectively. The changes in the parameters characterizing the magnetic hysteresis reflect the changes in the domain structure of the MnAs clusters, determined by their magnetic disorder and geometrical parameters.

\section{Magnetotransport data}

\begin{figure}
 \includegraphics[width = 0.45\textwidth, bb = 10 20 602 530]
{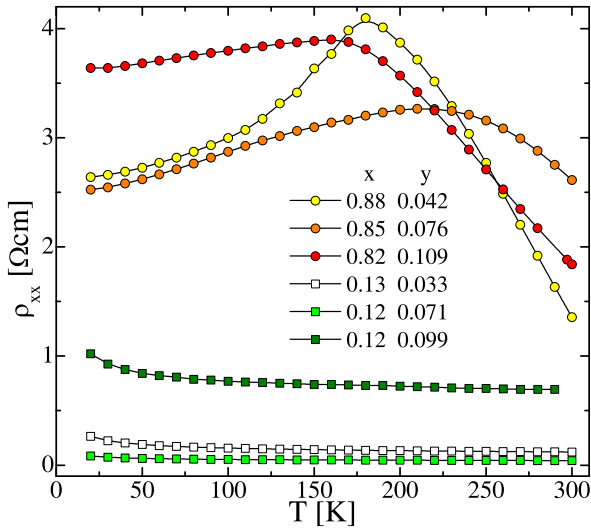}%
 \caption{\label{FigRT} Resistivity $\rho_{xx}$ as a function of temperature obtained for the Zn$_{1\textrm{-}x\textrm{-}y}$Cd$_{x}$Mn$_{y}$GeAs$_{2}$ crystals with different chemical compositions.}
 \end{figure}

\begin{figure}
 \includegraphics[width = 0.5\textwidth, bb = 10 10 540 610]
{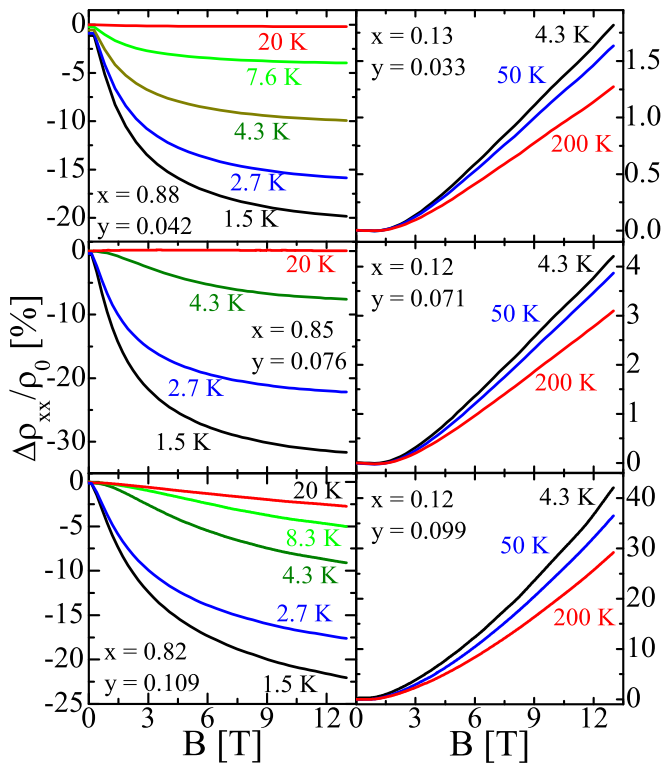}%
 \caption{\label{FigMRes} The magnetic field dependence of the resistivity $\rho_{xx}$ obtained at selected stabilized temperatures for the Zn$_{1\textrm{-}x\textrm{-}y}$Cd$_{x}$Mn$_{y}$GeAs$_{2}$ crystals with different chemical compositions.}
 \end{figure}

\begin{figure}
 \includegraphics[width = 0.45\textwidth, bb = 10 10 600 530]
{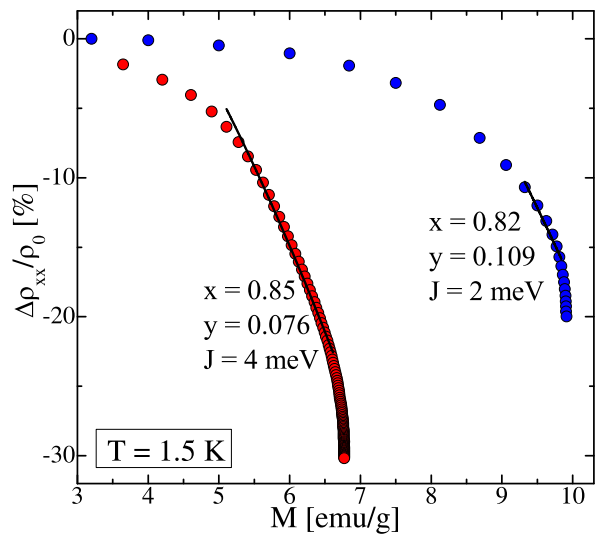}%
 \caption{\label{FigMRScal01} Magnetoresistance as a function of the magnetization obtained experimentally (points) and fitted using Eq.$\;$\ref{EqMr02} for two Zn$_{1\textrm{-}x\textrm{-}y}$Cd$_{x}$Mn$_{y}$GeAs$_{2}$ crystals with different chemical compositions (lines).}
 \end{figure}

\begin{figure}
 \includegraphics[width = 0.45\textwidth, bb = 10 10 600 530]
{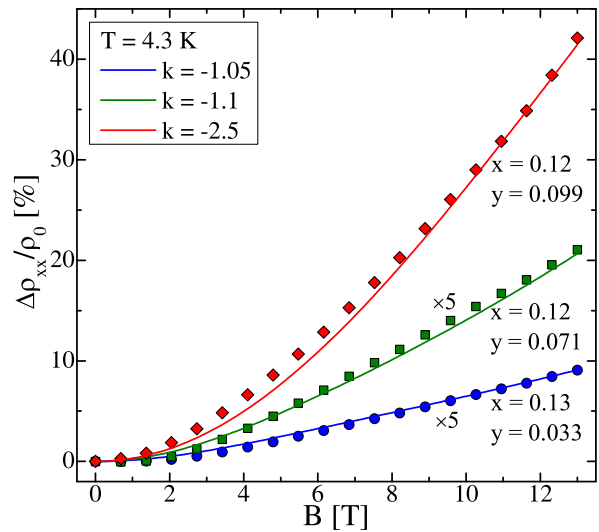}%
 \caption{\label{FigMRScal02} Magnetoresistance as a function of the magnetic field obtained experimentally (points) and fitted using Eq.$\;$\ref{EqLinMR01} (lines) for the selected Zn$_{1\textrm{-}x\textrm{-}y}$Cd$_{x}$Mn$_{y}$GeAs$_{2}$ crystals with different chemical compositions.}
 \end{figure}

The electrical properties of the Zn$_{1\textrm{-}x\textrm{-}y}$Cd$_{x}$Mn$_{y}$GeAs$_{2}$ crystals were studied by means of both temperature and field dependent magnetotransport measurements. We have used the superconducting magnet with maximum magnetic field equal to $B$$\,$$=$$\,$13$\;$T and a sweep speed of about 0.5$\;$T/min, equipped with the cryostat allowing the control of the temperature of the sample in the range of 1.4$\,$$\leq$$\,$$T$$\,$$\leq$$\,$300$\;$K. The samples, cut to the size of about 1$\times$1$\times$10$\;$mm, were etched and cleaned before making electrical contacts. The contacts were made with the use of gold wire and indium solder. The ohmic behavior of each contact pair was checked prior to proper measurements. The magnetoresistance and the Hall effect were measured simultaneously at selected temperatures with the use of the six contact dc current technique. \\ \indent Initially, we measured the temperature dependence of the resistivity component parallel to the current direction, $\rho_{xx}$, in the absence of an external magnetic field. The results of these measurements are plotted in Fig.$\;$\ref{FigRT}.
The $\rho_{xx}$($T$) for the group of samples with $x$$\,$$\approx$$\,$0.12 is a decreasing function of temperature while for the samples with $x$$\,$$\approx$$\,$0.85 the $\rho_{xx}$($T$) increases with $T$ at low temperatures, reaches a maximum and then is a decreasing function of temperature. The decrease of $\rho_{xx}$($T$) with $T$ which was observed for a group of samples with $x$$\,$$\approx$$\,$0.12 cannot be interpreted as thermally-activated transport because the conductivity $\sigma_{xx}$($T$) does not scale with $\sigma_{0}$$\cdot$$\exp(-E_{a}/k_{B}T)$, where $\sigma_{0}$ is the lattice conductivity, $E_{a}$ is the activation energy, and $k_{B}$ is the Boltzmann constant. The observed decrease of the $\rho_{xx}$($T$) in the samples with $x$$\,$$\approx$$\,$0.12 most probably originates from the existence of a large number of defects opening a second, metallic conduction channel and increasing the free carrier concentration of the alloy. A more complex $\rho_{xx}$($T$) dependence is observed for the samples with $x$$\,$$\approx$$\,$0.85. It is highly probable that the observed $\rho_{xx}$($T$) originated from the presence of large structural disorder in the samples. A better understanding of the observed transport phenomena requires the Hall effect results analysis. \\ \indent The Hall effect measured as a function of temperature allows us to determine the temperature dependence of the Hall carrier concentration $n$$\,$$=$$\,$$(eR_{H})^{-1}$, where $e$ is the elementary charge and $R_{H}$ is the Hall constant, and the Hall carrier mobility $\mu$$\,$$=$$\,$$(\rho_{xx} \cdot e \cdot n)^{-1}$ for all our samples at stabilized temperatures. The obtained results of the Hall effect measurements indicated that for all our samples we observed almost no temperature dependence of both $n$ and $\mu$. This indicates that the carrier transport in the studied samples is not due to thermal activation of band carriers. It is evident, that defect states have major influence on the conductivity of this alloy. The Hall effect results obtained at $T$$\,$$=$$\,$300$\;$K are gathered in Table$\;$\ref{tab:TranspParams1}.
\begin{table}[t]
\caption{\label{tab:TranspParams1}
Transport parameters obtained at room temperature for the Zn$_{1\textrm{-}x\textrm{-}y}$Cd$_{x}$Mn$_{y}$GeAs$_{2}$ crystals with different chemical compositions.}
\begin{tabular}{cccc}
\hline \hline \smallskip
 $x$ & $y$ & $n$$\pm$$\Delta$$n$  & $\mu$$\pm$$\Delta$$\mu$   \\
 & & [cm$^{-3}$] & cm$^{2}$/(V$\cdot$s)  \\ \hline
 0.88 & 0.042 & (-8.1$\pm$0.4)$\times$10$^{17}$ & 11$\pm$1      \\
 0.85 & 0.076 & (-1.1$\pm$0.1)$\times$10$^{18}$ & 2.5$\pm$0.3   \\
 0.82 & 0.109 & (-1.9$\pm$0.1)$\times$10$^{18}$ & 2.0$\pm$0.2   \\ \hline
 0.13 & 0.033 & (1.1$\pm$0.1)$\times$10$^{20}$  & 1.0$\pm$0.5   \\
 0.12 & 0.071 & (8.2$\pm$0.5)$\times$10$^{19}$  & 2.0$\pm$0.2   \\
 0.12 & 0.099 & (5.5$\pm$0.4)$\times$10$^{19}$  & 1.1$\pm$0.1   \\  \hline \hline
\end{tabular}
\end{table}
As we can see the Hall carrier concentration changes sign from n-type in the samples with $x$$\,$$\approx$$\,$0.85 to p-type in the samples with $x$$\,$$\approx$$\,$0.12. It is a signature, that the Cd-content changes the dominant electrically active defect types (defects that can be the source of conducting electrons or holes) in this alloy. Most probably at least several types of defects are present in the material. It is confirmed by the low values of $\mu$ at room temperature and the lack of temperature dependence of the mobility in all our samples. \\ \indent The magnetic field dependence of the resistivity parallel to the current direction, $\rho_{xx}$, namely magnetoresistance (MR), was measured in parallel to the Hall effect measurements. The measurements were performed over the temperature range from 1.4$\;$K to 200$\;$K. For easier data presentation the MR curves are presented in the form of $\Delta \rho_{xx}/ \rho_{xx}(0)$$\,$$=$$\,$$(\rho_{xx}(B)$$\,$$-$$\,$$\rho_{xx}(B=0))$/$\rho_{xx}(B=0)$. The selected results obtained for our Zn$_{1\textrm{-}x\textrm{-}y}$Cd$_{x}$Mn$_{y}$GeAs$_{2}$ are presented in Fig.$\;$\ref{FigMRes}.
MR of the two sets of samples having different Cd contents shows different shapes. For the samples with $x$$\,$$\approx$$\,$0.85 the MR shows negative values at $T$$\,$$<$$\,$20$\;$K and classical positive MR values proportional to $B^{2}$ at $T$$\,$$\geq$$\,$20$\;$K. Negative MR shows a decrease of the amplitude with increasing temperature. The maximum value of the negative MR is around -30\% at $B$$\,$$=$$\,$13$\;$T for the sample with $y$$\,$$=$$\,$0.076. However, the MR for none of our samples could be saturated with the maximum magnetic field used in our experiments. The shape of the MR curves shows that most probably the magnetic field needed to saturate the MR increases with the addition of Mn to the samples. Since the magnetization of the samples does show saturation at a magnetic field of about 3$\;$T it is evident that the MR in the alloy is not fully related to the presence of MnAs clusters. The MR observed in the group of samples with high Cd content, $x$$\,$$\approx$$\,$0.85, shows a behavior similar to that observed for ZnGeAs$_{2}$:MnAs samples\cite{Kilanski2010a} while for the group of samples with low Cd content, $x$$\,$$\approx$$\,$0.12, the MR shows a behavior similar to that observed for CdGeAs$_{2}$:MnAs samples.\cite{Kilanski2011b} It is therefore possible that the physical mechanism leading to MR in these alloys is related not only to the cation type. \\ \indent There are several different mechanisms that could lead to negative MR in the alloy. The Moriya-Kawabata spin-fluctuation theory\cite{Moriya1973a} predicts that the negative magnetoresistance should scale with $m$$\,$$=$$\,$1 or 2 for weakly or nearly ferromagnetic metals, respectively. The scaling analysis of MR to the $\Delta \rho_{xx}$/$\rho_{0}$$\,$$\propto$$\,$$B^{m}$ proportionality, where $m$ is the scaling factor, done for the results obtained at temperatures lower than 20$\;$K shows, that the MR results can be fitted with the value of the scaling factor $m$ around 0.7$\pm$0.1. Since the $m$ value for our samples is different from 1 or 2 we can exclude the spin-fluctuations from being the main physical mechanism responsible for the observed negative MR. Moreover, the dependence of the MR on the magnetization $M$ normalized to the saturation magnetization $M_{S}$, i.e. $M$/$M_{S}$ is neither square nor cubic. It is another signature that the negative MR for our Zn$_{1\textrm{-}x}$Mn$_{x}$GeAs$_{2}$ samples originates from processes that are not strictly related with magnetic impurity scattering. \\ \indent Negative MR has been reported in many inhomogeneous systems like Ni-SiO$_{2}$ alloys\cite{Gerber1997a} and is usually treated in the framework of the theory of spin-polarized electrons. The negative MR has been reported in numerous granular systems and is found to possess large values around 60\% (Ref.$\;$\onlinecite{Kilanski2010a}) as well as small values around 1\% (Ref.$\;$\onlinecite{Helman1976a}). The resistivity $\rho_{xx}$($B$,$T$) of a metal with ferromagnetic clusters can be expressed within the molecular field theory\cite{Helman1976a} using the approximate spin correlation function $m$ in the following form:
\begin{equation}\label{EqMr01}
   \frac{ \langle \overrightarrow{S_{1}} \bullet \overrightarrow{S_{2}} \rangle}{S^{2}} = m^{2}(B,T),
\end{equation}
where the momenta of two separate grains with spin momentum values, $\overrightarrow{S_{1}}$ and $\overrightarrow{S_{2}}$, have the same magnitude of the spin momentum equal to $S$. The relative magnetoresistance $\Delta \rho_{xx}$/$\rho_{0}$ can be expressed with the use of the following equation:
\begin{equation}\label{EqMr02}
    \Delta \rho_{xx}/\rho_{0} = -\frac{JP}{4k_{B}T} \big{[}m^{2}(B,T) - m^{2}(B = 0, T)\big{]},
\end{equation}
where $P$ is the polarization of the tunneling electrons, $k_{B}$ is the Boltzmann constant, and $J$ is the electronic exchange coupling constant within the ferromagnetic grains. The model described by Helman and Abeles,\cite{Helman1976a} Eqs.$\;$\ref{EqMr01} and \ref{EqMr02} above, reproduces the normalized magnetoresistance with only the intra-granular exchange constant $J$ and the carrier polarization $P$ as the fitting parameters, with the magnetization of the sample measured at the same temperature as the magnetoresistance. We used the $P$ value obtained for MnAs clusters equal to 45\% (value taken from the work by Panguluri et al., Ref.$\;$\onlinecite{Panguluri2003a}) in order to reduce the number of fitting parameters to $J$ only. The results of the fitting procedure are presented in Fig.$\;$\ref{FigMRScal01}.
It is evident that the MR cannot be reproduced well in the entire magnetic field range with the use of a model given by Eqs.$\;$\ref{EqMr01} and \ref{EqMr02}. It is a signature of the presence of additional physical mechanisms responsible for the MR curves at low temperatures. It is highly probable, that the weak localization phenomena, observed for homogeneous Zn$_{1\textrm{-}x}$Mn$_{x}$GeAs$_{2}$ samples with $x$$\,$$<$$\,$0.043 in Ref.$\;$\onlinecite{Kilanski2013a}, may occur at low fields in our samples. Moreover, at higher fields the classical positive MR can influence the data. However, the fits allow us to estimate the inter-grain exchange constant between the MnAs clusters embedded in the semiconductor matrix. The obtained values increase with the amount of Mn in the samples from $J$$\,$$=$$\,$1$\;$meV for $y$$\,$$=$$\,$0.042 to 4$\;$meV for $y$$\,$$=$$\,$0.109. The obtained $J$ values are more than an order of magnitude higher than the values reported in Ref.$\;$\onlinecite{Kilanski2010a} for Zn$_{1\textrm{-}x}$Mn$_{x}$GeAs$_{2}$ with $J$$\,$$<$$\,$0.1$\;$meV. However, the higher values of $J$ correspond to the higher temperatures at which the negative MR is observed in our samples. \\ \indent For the samples with $x$$\,$$\approx$$\,$0.12 the MR has positive values and over the magnetic field range above 3$\;$T is either nearly linear with the magnetic field for $y$$\,$$=$$\,$0.033 and $y$$\,$$=$$\,$0.071 or nearly quadratic with the magnetic field for $y$$\,$$=$$\,$0.099. The MR amplitude observed at $T$$\,$$=$$\,$4.3$\;$K increases with the amount of Mn in the samples from about 1.5\% for the samples with $y$$\,$$=$$\,$0.033 up to about 40\% for the samples with $y$$\,$$=$$\,$0.099. The positive MR is observed at temperatures up to 200$\;$K and shows a slow decrease with increasing temperature. Since most of Mn ions in our samples are clustered into MnAs inclusions it is obvious that the increase of $y$ leads to an increase of the concentration of the MnAs clusters. It is therefore possible, that the MR observed for our samples originates from the same physical mechanism as the MR reported for Cd$_{1\textrm{-}x}$Mn$_{x}$GeAs$_{2}$ in Ref.$\;$\onlinecite{Kilanski2011b}. \\ \indent The presence of a positive linear magnetoresistance is commonly attributed to the presence in the host material of inhomogeneities, having the conductivity different from the homogeneous host material. The linear, cluster-mediated magnetoresistance can have either large values of the order of 1000\% in MnAs-GaAs\cite{Johnson2010a} or values smaller than 1\% in case of thin Ni layers.\cite{Gerber1997a} An effective medium approximation (EMA) model describes quantitatively the classical geometrical magnetoresistance of inhomogeneous media.\cite{Guttal2005a, Guttal2006a} The model assumes a fixed balance between phases A and B, called $p_{A}$ and $p_{B}$$\,$$=$$\,$1$-$$p_{A}$, respectively, and defined zero field conductivities $\sigma_{0A}$ and $\sigma_{0B}$. The effective conductivity of the material, $\sigma_{eff}$, can be calculated using the coupled self consistent equations, which can be expressed in a matrix form. The magnetic field dependence of resistivity can be calculated by solving the self consistent equation
\begin{equation}\label{EqLinMR01}
    \sum_{i=A,B} = p_{i} \delta \sigma_{i}(I-\Gamma\sigma_{i})^{-1} = 0,
\end{equation}
where $\sigma_{i}$ is the conductivity of either $A$ or $B$ phase, $\delta \sigma_{i}$$\,$$=$$\,$$\sigma_{i}$$-$$\sigma$, $\sigma$ is the experimental macroscopic conductivity, $\Gamma$ is the depolarization tensor describing the geometrical properties of clusters and $I$ is the unity matrix. The EMA model\cite{Guttal2005a, Guttal2006a} is used to reproduce the experimental MR curves with given carrier conductivity of the host matrix, $\sigma_{A}$ (taken from low temperature Hall measurements) and the ratio between the carrier Hall constants $R_{HA}$ and $R_{HB}$ in the two phases A and B given by the parameter $k$ equal to $k$$\,$$=$$\,$$R_{HA}$/$R_{HB}$ being the fitting parameter. The MR curves calculated using the EMA model are presented together with experimental data obtained at $T$$\,$$=$$\,$4.5$\;$K for the selected Zn$_{1\textrm{-}x\textrm{-}y}$Cd$_{x}$Mn$_{y}$GeAs$_{2}$ samples having different chemical compositions in Fig.$\;$\ref{FigMRScal02}. We obtain good agreement between the theoretical and experimental curves in the case of all the samples under investigation. The best fits to the experimental data are obtained for the negative $k$ values indicating that the clusters present in the semiconductor matrix have the opposite ($n$-type) conductivity type.
The absolute value of $k$ increases as a function of the Mn content, $y$, indicating an increasing difference between the carrier concentration of the host lattice and MnAs inclusions.

\section{Summary}

We explored the structural, magnetic, and electrical properties of Zn$_{1\textrm{-}x\textrm{-}y}$Cd$_{x}$Mn$_{y}$GeAs$_{2}$  nanocomposite ferromagnetic semiconductor samples with Cd-content $x$$\,$$\approx$$\,$0.12 or 0.85 and changeable Mn content, $y$, ranging from 0.033 to 0.109. The presence of MnAs clusters induces room temperature ferromagnetism with the Curie temperature, $T_{C}$, around 305$\;$K in all our samples. The magnetization $M$($B$) curves for the samples with $x$$\,$$\approx$$\,$0.12 reach saturation at $B$$\,$$<$$\,$9$\;$T but do not show magnetization hysteresis, while for the samples with $x$$\,$$\approx$$\,$0.85 the $M$($B$) curves do not saturate even at $B$$\,$$=$$\,$9$\;$T and show well defined hysteresis loops with the maximum coercive field $H_{C}$ equal to 28$\;$mT for the Zn$_{0.071}$Cd$_{0.82}$Mn$_{0.109}$GeAs$_{2}$ sample. \\ \indent The magnetotransport of the studied alloy is influenced by both Cd and Mn content. The sign of the MR changes with the Cd content, $x$, from negative occurring only at $T$$\,$$<$$\,$20$\;$K observed for the samples with $x$$\,$$\approx$$\,$0.85 to positive MR sign observed for the samples with $x$$\,$$\approx$$\,$0.12. Three major mechanisms cause the negative MR of the samples with $x$$\,$$\approx$$\,$0.85: carrier polarization by granular ferromagnetic clusters, weak localization due to defects, and the classical orbital MR. On the other hand the geometrical MR is the main physical mechanism responsible for the positive MR observed for the samples with $x$$\,$$\approx$$\,$0.12.

\section{Acknowledgments}

\noindent Scientific work was financed from funds for science in 2011-2014, under the project no.  N202 166840 granted by the National Center for Science of Poland. This work has been supported by the RFBR Project No. 13-03-00125 and Russian Federation Project No. MK-1454.2014.3.

\end{document}